\documentclass[11pt,onecolumn]{article}
\usepackage[a4paper, top=25mm, bottom=25mm, left=20mm, right=20mm]{geometry}

\usepackage{graphicx}
\usepackage{amsmath}
\usepackage{amssymb}
\usepackage{url}
\usepackage{subcaption}

\usepackage[dvipsnames]{xcolor}
\usepackage[pagebackref=true,breaklinks=true,colorlinks,bookmarks=false]{hyperref}
\usepackage{microtype}
\title{Physics-Informed Neural Networks for the Time-Domain Maxwell Equations with Split-Field Perfectly Matched Layers}

\author{
Xiaodong Liu\\
Remcom Inc, State College, PA 16801, USA\\
\texttt{liufield@gmail.com}
\and
Lingquan Li\\
Shanghai University, Shanghai 200072, China
\and
Gregory Moss\\
Remcom Inc, State College, PA 16801, USA
\and
Scott Langdon\\
Remcom Inc, State College, PA 16801, USA
}

\date{\small Preprint. Submitted to IEEE AP-S 2026 Meeting.}

\begin{document}
\maketitle
\begin{abstract}
Physics-informed neural networks (PINNs) incorporate Maxwell’s equations, initial conditions, boundary conditions, and measurement data directly into the learning process, transforming the solution of partial differential equations into a constrained optimization problem. As such, PINNs are attracting increased attention in computational electromagnetics as an alternative to traditional time-domain solvers.

This paper presents a PINN formulation for the time-domain Maxwell's equations incorporating split-field perfectly matched layers (PMLs). A key advantage of the split-field PML formulation is that the same governing equations can be applied in both the physical and PML regions, simplifying the PINN formulation and loss construction.

The proposed approach is validated using one-dimensional and two-dimensional Gaussian pulse problems with PML. The PINN solutions show good agreement with analytical and finite-difference time-domain (FDTD) reference solutions, demonstrating the feasibility of combining PINNs with split-field PMLs for open-domain time-domain electromagnetic simulations.
\end{abstract}

\section{Introduction}
Unlike purely data-driven models, physics-informed neural networks (PINNs), originally proposed by M. Raissi \cite{article:RaissiPINNs}, encode the underlying PDEs, initial conditions, and boundary conditions into the learning process, thereby enabling mesh-free and data-efficient solution strategies. This paradigm is becoming increasingly popular for solving forward and inverse problems in computational physics, including fluid dynamics, acoustics and electromagnetics. Significant progress has been made using PINNs, including strict boundary condition enforcement,  random weight factorization \cite{article:WangRWF}, effective optimization algorithms \cite{article:NocedalBFGS, article:RathoreOptimize, article:OrenSSVM}, and causality-enhanced strategies \cite{article:WangCausality,article:LiDisccretePINNs}.

The time-domain Maxwell's equations govern the evolution of electric and magnetic fields and form the foundation for modeling wave propagation, scattering, and radiation phenomena. Numerical methods such as the finite-difference time-domain (FDTD) method \cite{article:YeeFDTD} and finite element method (FEM) \cite{article:JinFEMbook} have achieved great success; however, they often require explicit meshing and careful treatment of complex geometries. PINNs offer an alternative approach by approximating the electromagnetic fields with a neural network and enforcing Maxwell’s equations, initial conditions, and boundary conditions as constraints in the loss function, which naturally avoids explicit meshing. 

To simulate open-domain electromagnetic problems, artificial absorbing boundaries are required to truncate the computational domain while minimizing spurious reflections. The perfectly matched layer (PML) \cite{article:BerengerSplitPML, article:GedneyPML} remains one of the most effective absorbing boundary techniques. The split-field PML formulation achieves reflectionless absorption by decomposing field components and introducing auxiliary damping terms within the PML region. PML has been widely adopted in FDTD and FEM due to its robustness and simplicity of implementation.

Despite its effectiveness in conventional solvers\cite{article:PINNsTimeMaxwell, article:ZhengImplementationPINN,
article:NohraPINNsdiscontinuousmedia, article:ShavinerPINNstimeMaxwell}, the incorporation of PML into PINNs has not been widely explored. In this work, we develop a PINN framework for solving the time-domain Maxwell equations with split-field PML. The proposed approach embeds both the Maxwell equations in the physical domain and the split-field PML equations in the absorbing region into a unified loss functional. By treating all the electromagnetic fields as network outputs, the method provides a physically consistent, mesh-free solver for open-domain electromagnetic wave propagation.

\section{PINNs for Time-Domain Maxwell's Equations}

\subsection{Time-Domain Split-Field PML Equations}
In the two-dimensional TM$_z$ mode, the electromagnetic fields have only three nonzero components: the out-of-plane electric field $E_z$ and the in-plane magnetic fields $H_x$ and $H_y$. By implementing $E_z = E_{zx} + E_{zy} $, the time-domain split-field PML equations reduce to
\begin{equation}
\begin{aligned}
{\varepsilon_r}\frac{\partial E_{zx}}{\partial t} +\sigma_x E_{zx}&=  
\frac{\partial H_y}{\partial x} , \\
{\varepsilon_r}\frac{\partial E_{zy}}{\partial t} +\sigma_y E_{zy} &=  
- \frac{\partial H_x}{\partial y}, \\
{\mu_r}\frac{\partial H_x}{\partial t} + \sigma_y^*H_x&= - \frac{\partial E_z}{\partial y}, \\
{\mu_r}\frac{\partial H_y}{\partial t} + \sigma_x^*H_y&=  \frac{\partial E_z}{\partial x}.
\end{aligned}
\label{eq:2D}
\end{equation}
where $\varepsilon_r$, $\mu_r$, $\sigma$ and $\sigma^*$ denote the relative electric permittivity, relative magnetic permeability, electric conductivity, and magnetic loss of the medium, respectively. The above equations are nondimensional and are used in the PINN formulation to accelerate convergence. The reference values are listed in Table \ref{Table1Label}.

\begin{table}[htbp]
\begin{center}
\caption{Reference physical scales for nondimensional time-domain Maxwell's equations} \label{Table1Label}
\begin{tabular}{|c|c|c|}
\hline
Quantity (unit) & Symbol & Typical choice \\
\hline
Length (m) & $L_0$ & Domain size \\
\hline
Time (s) & $T_0$ & $L_0 / c$ \\
\hline
Electric field (V/m)& $E_0$ & Peak field amplitude \\
\hline
Magnetic field (A/m)& $H_0$ & $E_0 / Z_0$ \\
\hline
Speed of light (m/s)& $c$ & $1/\sqrt{\mu_0\varepsilon_0}$ \\
\hline
Impedance ($\Omega$) & $Z_0$ & $\sqrt{\mu_0/\varepsilon_0}$ \\
\hline
Electric conductivity ($1/(\Omega m)$)& $\sigma_0$ & $1/({L_0Z_0})$ \\
\hline
Magnetic conductivity ($\Omega/m$) & $\sigma^*_0$ & $Z_0/L_0$ \\
\hline
\end{tabular}
\end{center}
\end{table}

If $\sigma_x=\sigma_y=0$ and $\sigma_x^*=\sigma_y^*=0$, the above equations reduce to Maxwell's equations in a lossless medium (e.g., vacuum). As a result, all computational domains, including the PML region, can be represented using a uniform set of equations. This is a desirable property for PINNs. The input and output dimensions remain the same across the physical and PML regions, thereby avoiding the additional complexity of handling convolutional terms required in other PML formulations \cite{article:GedneyPML}.

To ensure impedance matching between the physical domain and the PML region and to minimize spurious reflections at the PML interface, the electric and magnetic loss parameters are chosen to satisfy the standard PML impedance-matching condition,
\begin{equation}
\begin{aligned}
\frac{\sigma}{\varepsilon_r} = \frac{\sigma^*}{\mu_r}.
\end{aligned}
\label{eq:sigma}
\end{equation}
Taking $\sigma_x$ as an example, polynomial grading can be applied as
\begin{equation}
\begin{aligned}
\sigma_x = \sigma_{\max}(\frac{\Delta x}{d})^m,
\end{aligned}
\label{eq:sigmax}
\end{equation}
where $\Delta x$ denotes the distance from the PML interface, $d$ is the PML thickness, and $m$ is the grading order.
For 1D, $x$-directed, $z$-polarized TEM mode, the above equation reduces to \eqref{eq:1D}, 
\begin{equation}
\begin{aligned}
{\varepsilon_r}\frac{\partial E_{z}}{\partial t} +\sigma_x E_{z}&=  
\frac{\partial H_y}{\partial x} , \\
{\mu_r}\frac{\partial H_y}{\partial t} + \sigma_x^*H_y&=  \frac{\partial E_z}{\partial x}.
\end{aligned}
\label{eq:1D}
\end{equation}

\subsection{Loss Functions}
The PINNs take $(x, y, t)$ as input and employ a fully connected neural network to predict the electromagnetic field components $(E_{zx}, E_{zy}, H_x, H_y)$. The loss function $L$ consists of several terms: the physics-informed loss term $L_{\text{physics}}$, terms that evaluate the error between the values predicted by the network and the prescribed initial and boundary data, $L_{\text{initial}}$ and $L_{\text{boundary}}$, and a term for other measured data, $L_{\text{measure}}$.
\begin{equation}
L = \lambda_1 L_{\text{physics}} + \lambda_2 L_{\text{initial}} + \lambda_3 L_{\text{boundary}} + \lambda_4 L_{\text{measure}}.
\end{equation}
Here, the $\lambda$ values denote the weights for each term. Unless otherwise specified, $\lambda = 1$ for all terms.

The physics-informed term enforces the residuals of the governing PDEs at a set of collocation points $\{ \boldsymbol{x}_p^i, t_p^i \}_{i=1}^{N_p}$:
\begin{equation}
L_{\text{physics}} = \frac{1}{N_p} \sum_{i=1}^{N_p} \| \boldsymbol{R}(\boldsymbol{x}_p^i, t_p^i; \theta) \|^2,
\end{equation}
where $\boldsymbol{R}$ denotes the PDE residual vector ($\boldsymbol{R}$ has 4 and 2 components for \eqref{eq:2D} and \eqref{eq:1D}, respectively), and $\theta$ represents the neural network parameters.

The initial condition term penalizes deviations from the prescribed initial values at points $\{ \boldsymbol{x}_0^i \}_{i=1}^{N_0}$:
\begin{equation}
L_{\text{initial}} = \frac{1}{N_0} \sum_{i=1}^{N_0} \| \boldsymbol{u}(\boldsymbol{x}_0^i, 0; \theta) - \boldsymbol{u}_0(\boldsymbol{x}_0^i) \|^2,
\end{equation}
where $\boldsymbol{u}$ denotes the predicted field values and $\boldsymbol{u}_0$ denotes the prescribed initial values.

The boundary term enforces the network predictions at boundary points $\{ \boldsymbol{x}_b^i, t_b^i \}_{i=1}^{N_b}$:
\begin{equation}
L_{\text{boundary}} = \frac{1}{N_b} \sum_{i=1}^{N_b} \| \boldsymbol{u}(\boldsymbol{x}_b^i, t_b^i; \theta) - \boldsymbol{u}_b(\boldsymbol{x}_b^i, t_b^i) \|^2.
\end{equation}

If available, measurements at points $\{ \boldsymbol{x}_m^i, t_m^i \}_{i=1}^{N_m}$ can be incorporated:
\begin{equation}
L_{\text{measure}} = \frac{1}{N_m} \sum_{i=1}^{N_m} \| \boldsymbol{u}(\boldsymbol{x}_m^i, t_m^i; \theta) - \boldsymbol{u}_m^i \|^2,
\end{equation}
where $\boldsymbol{u}_m^i$ denotes the observed or measured field values.

Unless otherwise specified, Latin hypercube sampling (LHS) is adopted for data generation, and the hyperbolic tangent (\texttt{tanh}) function is used as the network activation. The resulting optimization problem is solved using a two-stage training strategy: the Adam optimizer \cite{article:Adam} is first employed for stochastic gradient-based pretraining, followed by the limited-memory Broyden--Fletcher--Goldfarb--Shanno (L-BFGS) algorithm \cite{article:LBFGS} to further refine the solution and enhance convergence. All implementations are carried out using the DeepXDE framework \cite{article:DeepXDE}.

\section{Numerical Examples}

\subsection{1D Gaussian Pulse}
The 1D problem is governed by \eqref{eq:1D}. The time-decaying Gaussian electric pulse is imposed as a Dirichlet boundary condition at the left boundary $z = 0$,
\begin{equation}
E_z(t) = e^{-\frac{(t - t_0)^2}{2\sigma^2}}.
\end{equation}
Here, $t_0 = 0.5$ and $\sigma = 0.05$.

The first test case examines the propagation of a Gaussian pulse in a vacuum. The domain $(x,t)$ is $[0, 1] \times [0, 1]$. The simulation time window is chosen such that the propagating pulse does not reach the boundary $x = 1.0$; therefore, PML is not required. For the PINN simulations, $N_p = 10000$, $N_0 = 500$, and $N_b = 500$. The excitation term induced by the Gaussian pulse is treated as measured time-dependent data with $N_m = 200$. These measured data are sampled at uniformly distributed time intervals at $x=0$. The same sampling methodology is applied to other cases, where applicable. The network consists of four hidden layers, each containing 40 neurons. 

\begin{figure}[!t]
\centering
\begin{subfigure}[b]{0.48\textwidth}
  \centering
  \includegraphics[width=\linewidth]{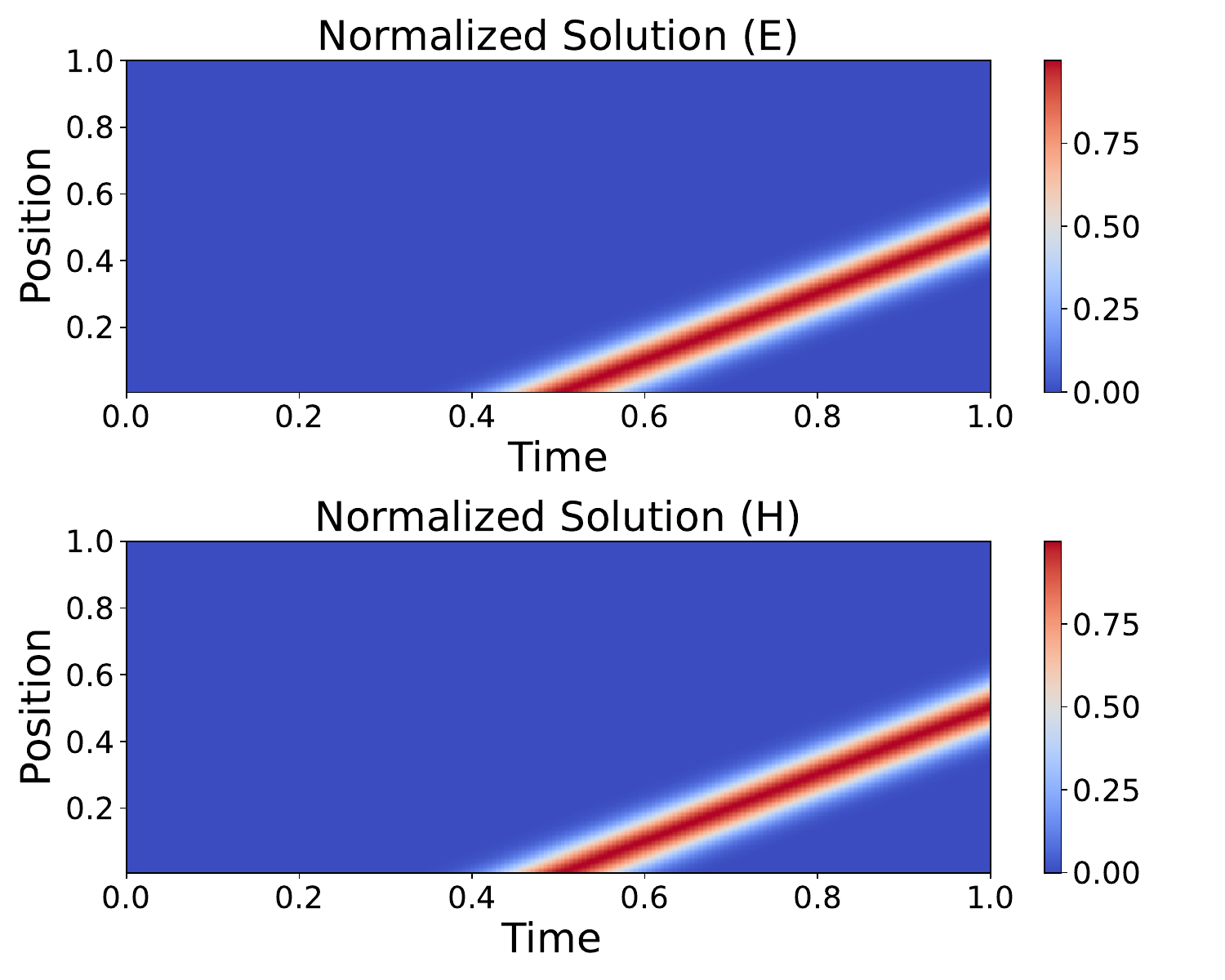}
  \caption{}
  \label{fig:EzField1DNoPML}
\end{subfigure}
\hfill
\begin{subfigure}[b]{0.48\textwidth}
  \centering
  \includegraphics[width=\linewidth]{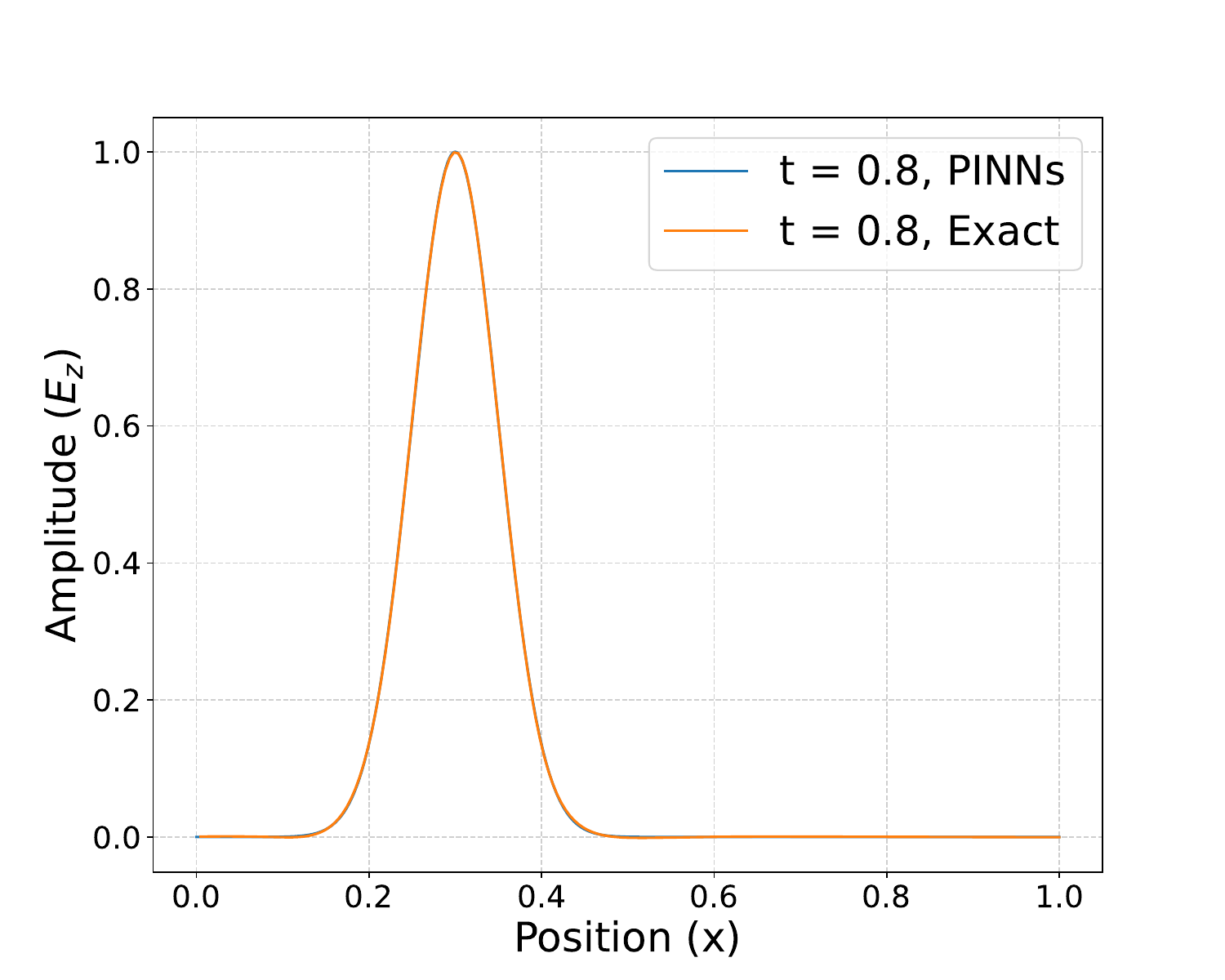}
  \caption{}
  \label{fig:EzComp1DNoPML}
\end{subfigure}
\caption{(a) Electric $E_z$ and magnetic $H_y$ fields. (b) Comparison between PINNs and the exact solution at $t = 0.8$.}
\label{Fig1Label}
\end{figure}

The electric and magnetic fields over the computational domain are presented in Fig.~\ref{fig:EzField1DNoPML}. The numerical results at $t = 0.8$ from PINNs are directly compared with the analytical solution in Fig.~\ref{fig:EzComp1DNoPML}, showing that PINNs are able to provide satisfactory results.

The second test case considers a Gaussian pulse propagating in a heterogeneous medium. The domain is $(x,t) \in [0, 1] \times [0, 1.3]$. A dielectric interface is located at $x = 0.5$, with the relative permittivity changing from $\varepsilon_r = 1$ to $\varepsilon_r = 4$. As in the first test, the simulation is time-windowed such that the pulse does not reach the outer boundary; therefore, eliminating the need for PML. Here, $N_p = 20000$, $N_0 = 500$,  $N_b = 500$, and  $N_m=325$. The network consists of four hidden layers, each containing 60 neurons.  

\begin{figure}[!ht]
\centering
\begin{subfigure}[b]{0.48\textwidth}
  \centering
  \includegraphics[width=\linewidth]{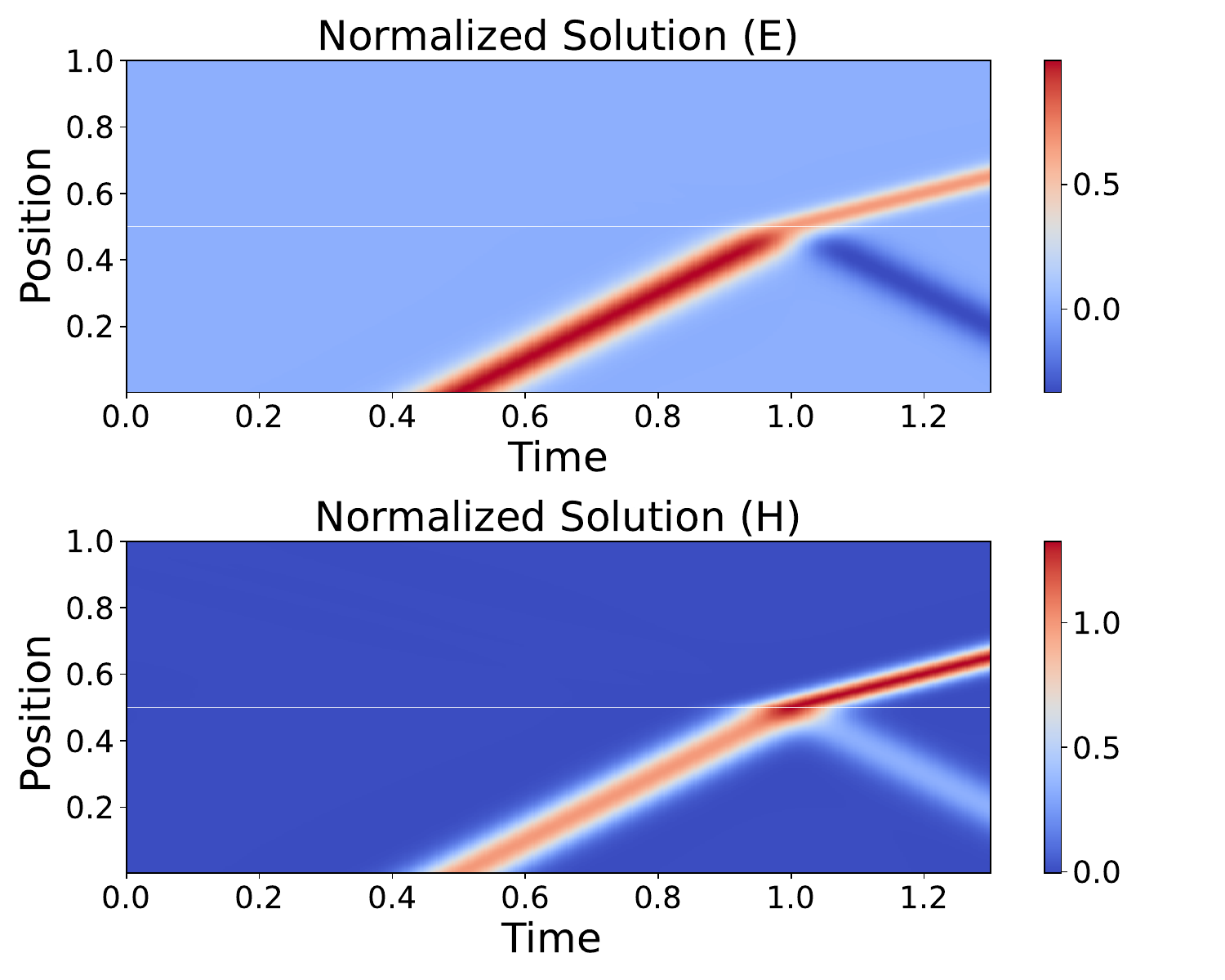}
  \caption{}
  \label{fig:EzField1DNoPML2Mediums}
\end{subfigure}
\hfill
\begin{subfigure}[b]{0.48\textwidth}
  \centering
  \includegraphics[width=\linewidth]{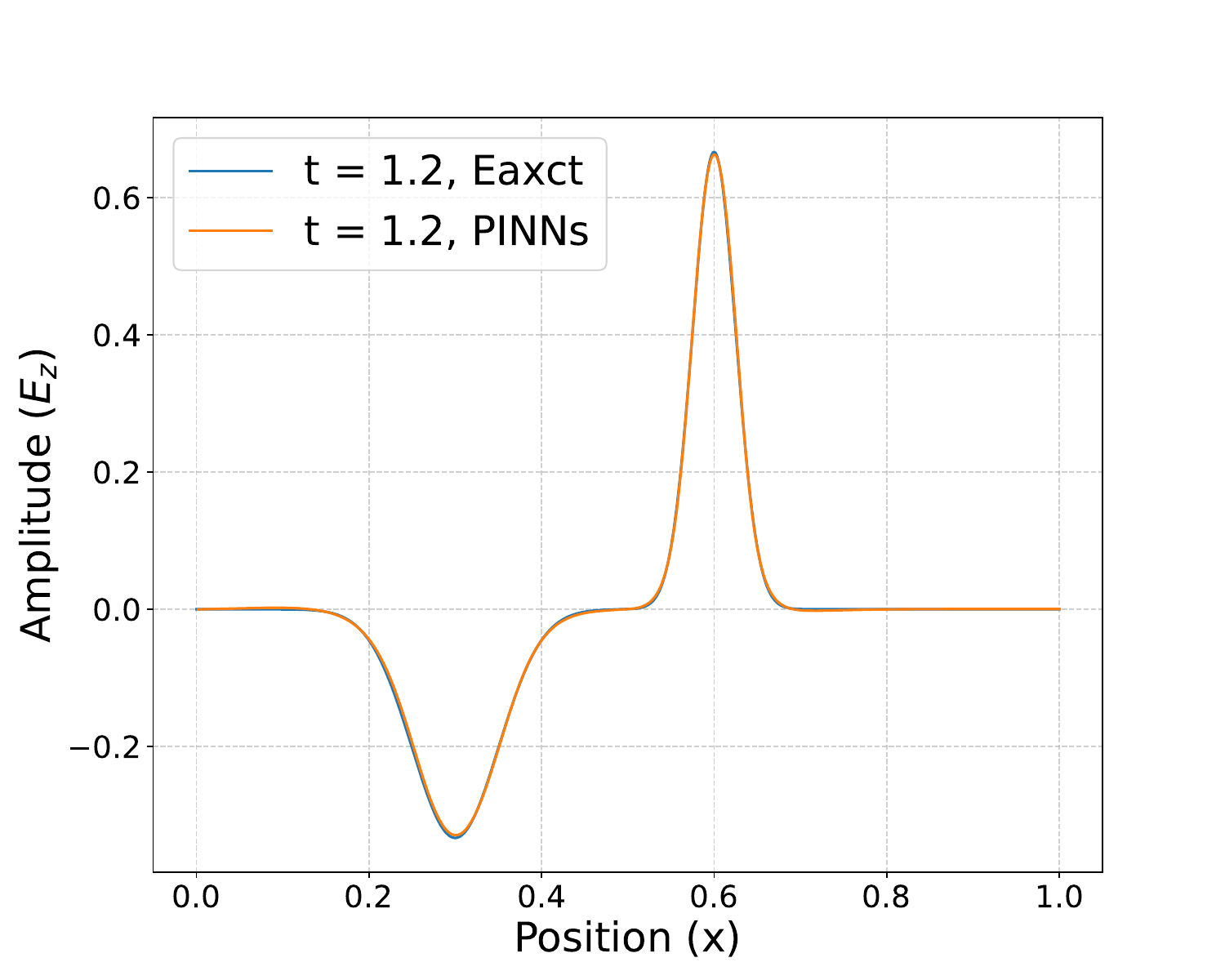}
  \caption{}
  \label{fig:EzComp1DNoPML2Mediums}
\end{subfigure}
\caption{(a) Electric $E_z$ and magnetic $H_y$ fields. (The white
line represent the interface at $x=0.5$) (b) Comparison between PINNs and the exact solution at $t = 1.2$.}
\label{Fig2Label}
\end{figure}

The electric and magnetic fields over the domain are presented in Fig.~\ref{fig:EzField1DNoPML2Mediums}. Due to the dielectric interface, the electromagnetic wave is partially reflected. The numerical results at $t = 1.2$ from PINNs are directly compared with the analytical solution in Fig.~\ref{fig:EzComp1DNoPML2Mediums}, showing that PINNs are able to provide satisfactory results.

The third case utilizes a PML region. The domain is $(x,t) \in [0, 1] \times [0, 1.5]$. For $x < 0.6$, the medium is vacuum with $\varepsilon_r = 1.0$; for $x > 0.6$, the region is modeled as PML with $\varepsilon_r = 1.0$. Here, $N_p = 15000$, $N_0 = 1000$, $N_b = 1000$  and $N_m = 150$. The network consists of four hidden layers, each containing 60 neurons. The PML parameters are $\sigma_{\max} = 40$ and $m = 3$.  

\begin{figure}[!ht]
\centering
\begin{subfigure}[b]{0.48\textwidth}
  \centering
  \includegraphics[width=\linewidth]{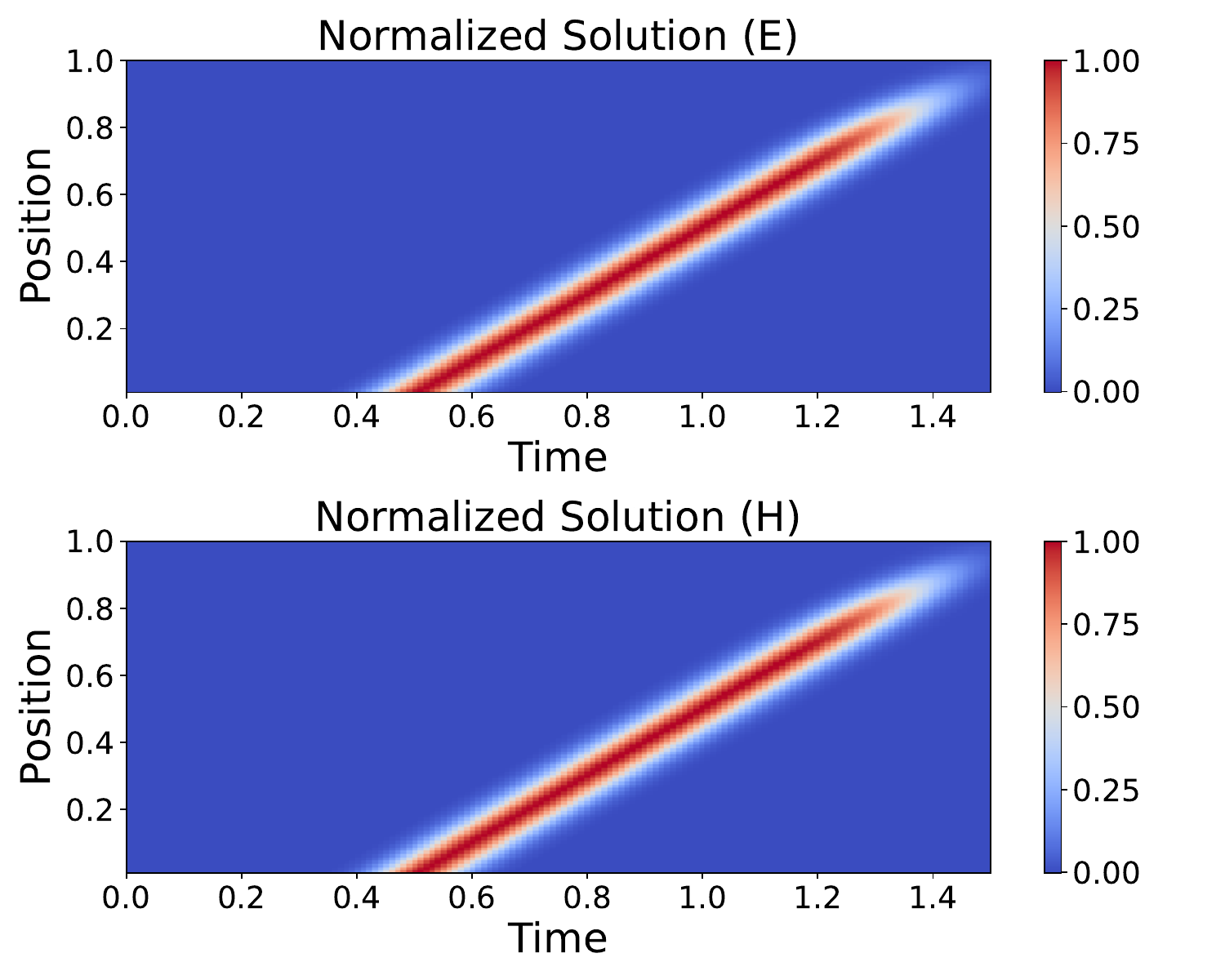}
  \caption{}
  \label{fig:EzField1DPML1Mediums}
\end{subfigure}
\hfill
\begin{subfigure}[b]{0.48\textwidth}
  \centering
  \includegraphics[width=\linewidth]{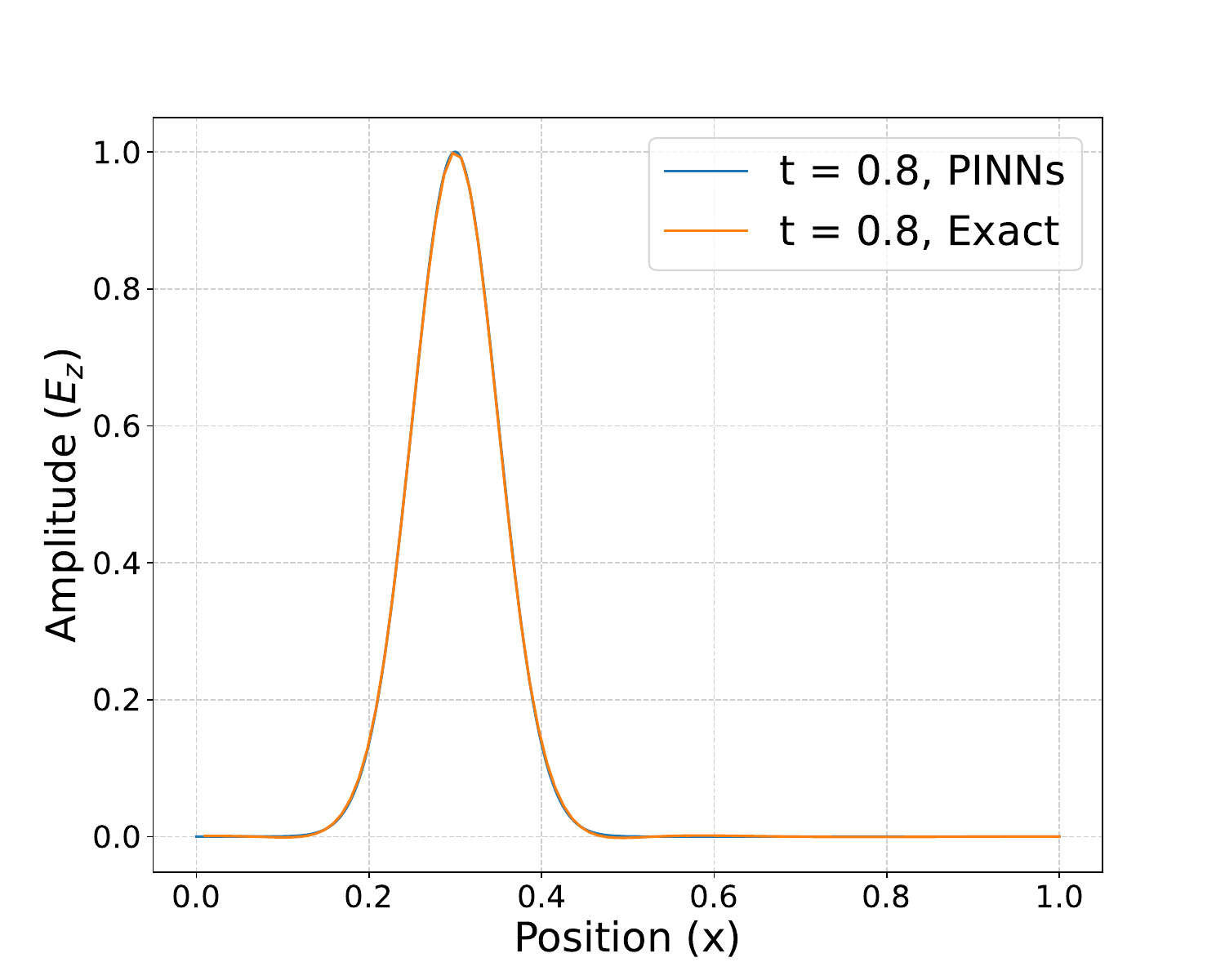}
  \caption{}
  \label{fig:EzComp1DPML1Mediums}
\end{subfigure}
\caption{(a) Electric $E_z$ and magnetic $H_y$ fields. (b) Comparison between PINNs and FDTD at $t = 0.8$.}
\end{figure}

The electric and magnetic fields are shown in Fig.~\ref{fig:EzField1DPML1Mediums}. The fields are dissipated through the PML region as expected. The $E_z$ distribution from PINNs matches well with the reference solution at $t = 0.8$, as shown in Fig.~\ref{fig:EzComp1DPML1Mediums}.

\subsection{2D Gaussian Pulse in Vacuum}
The final test case considers the propagation of a two-dimensional Gaussian pulse in vacuum. The computational space--time domain is $(x,y) \in [0.0, 1.0] \times [0.0, 1.0]$ with $t \in [0, 0.8]$. The PML interfaces are placed at $x = 0.2$, $x = 0.8$, $y = 0.2$, and $y = 0.8$. The PML parameters are $\sigma_{\max} = 10$ and $m = 3$. The initial condition consists of a Gaussian electric field pulse with zero initial magnetic field, given by
\begin{equation}
\begin{aligned}
E_{z}(x,y,t=0) &= e^{-40\left((x-0.5)^2 + (y-0.5)^2\right)}, \\
H_{x}(x,y,t=0) &= 0, \\
H_{y}(x,y,t=0) &= 0.
\end{aligned}
\label{eq:2D_IC}
\end{equation}

\begin{figure}[!ht]
\centering
\begin{subfigure}[b]{0.48\textwidth}
  \centering
  \includegraphics[width=\linewidth]{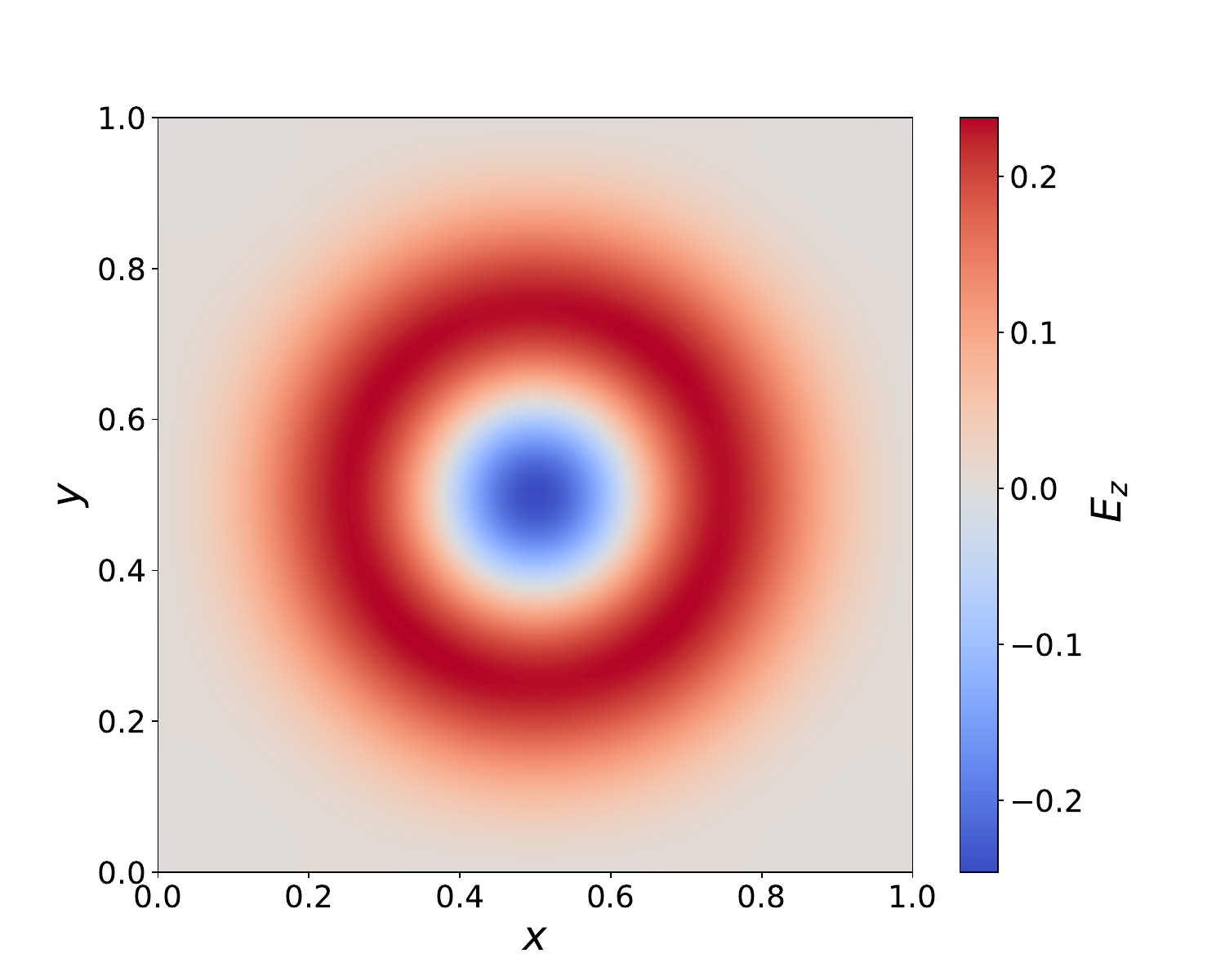}
  \caption{}
  \label{fig:Ez2DGaussianPML}
\end{subfigure}
\begin{subfigure}[b]{0.48\textwidth}
  \centering
  \includegraphics[width=\linewidth]{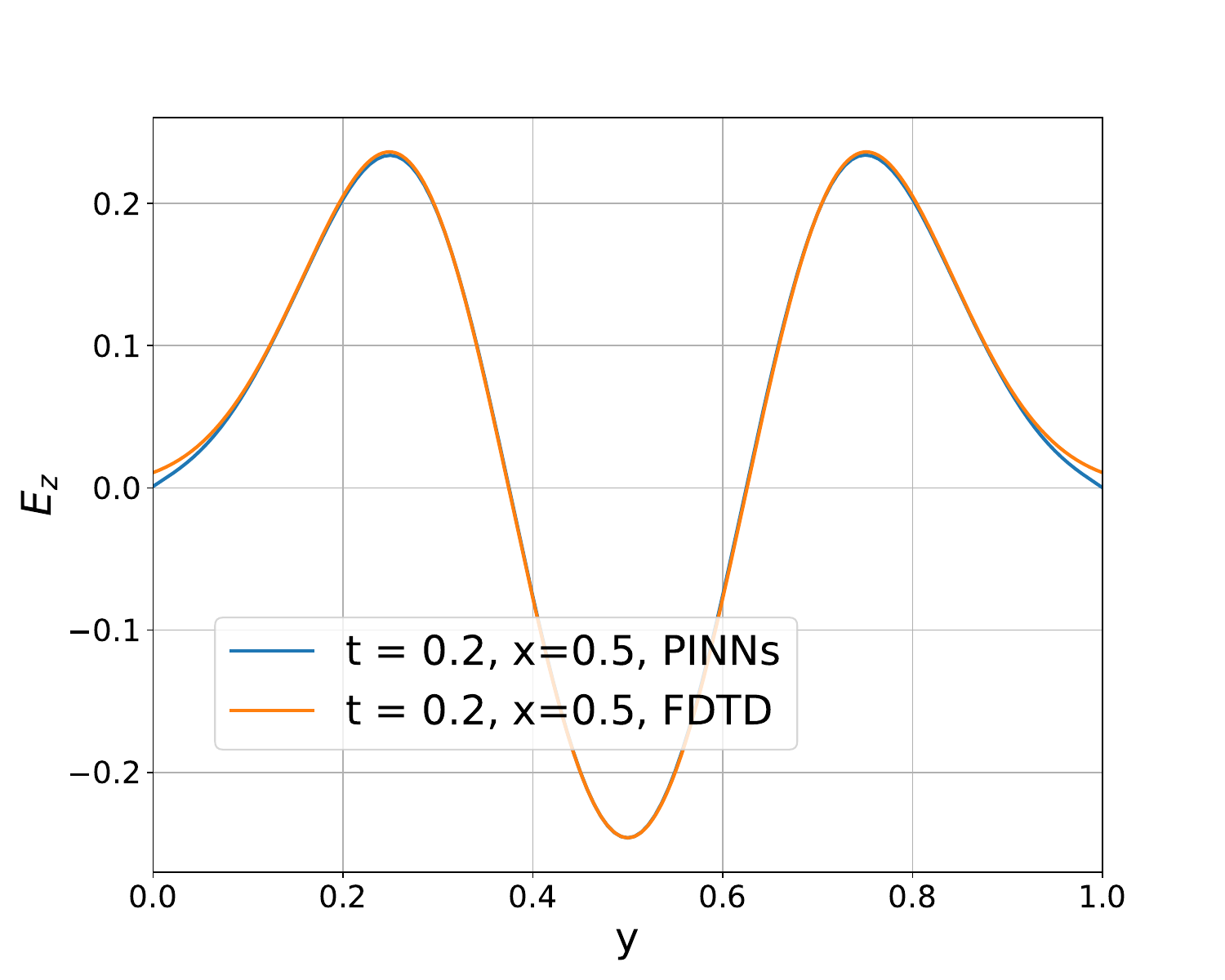}
  \caption{}
  \label{fig:EzComp2DGaussianPML}
\end{subfigure}
\caption{(a) Electric field $E_z$  at $t = 0.2$.  (b) The solution comparison between PINNs and FDTD  at $x=0.5$ and $t = 0.2$.}
\end{figure}

The $E_z$ field at $t = 0.2$ is shown in Fig.~\ref{fig:Ez2DGaussianPML}. An FDTD solution with grid spacing $\Delta x = \Delta y = 0.00125$ is used as the reference solution. Here, $N_p = 100000$, $N_0 = 1000$, and $N_b = 1000$. The network consists of four hidden layers, each containing 80 neurons. Sobol sequences are adopted for data generation and the sine function is used as the network activation.

\begin{figure}[t]
\centering
\begin{subfigure}[b]{0.48\textwidth}
  \centering
  \includegraphics[width=\linewidth]{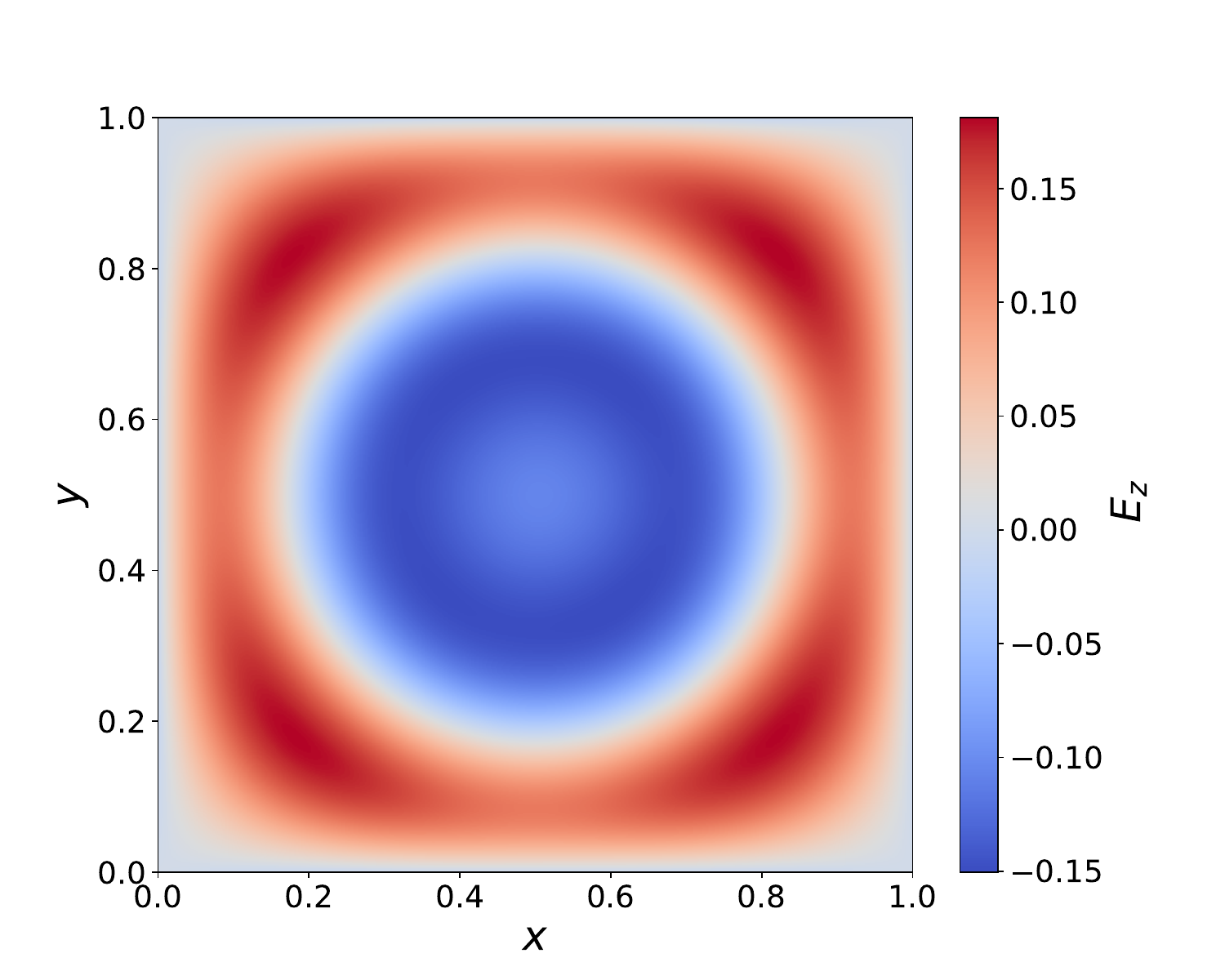}
  \caption{}
\end{subfigure}
\begin{subfigure}[b]{0.48\textwidth}
  \centering
  \includegraphics[width=\linewidth]{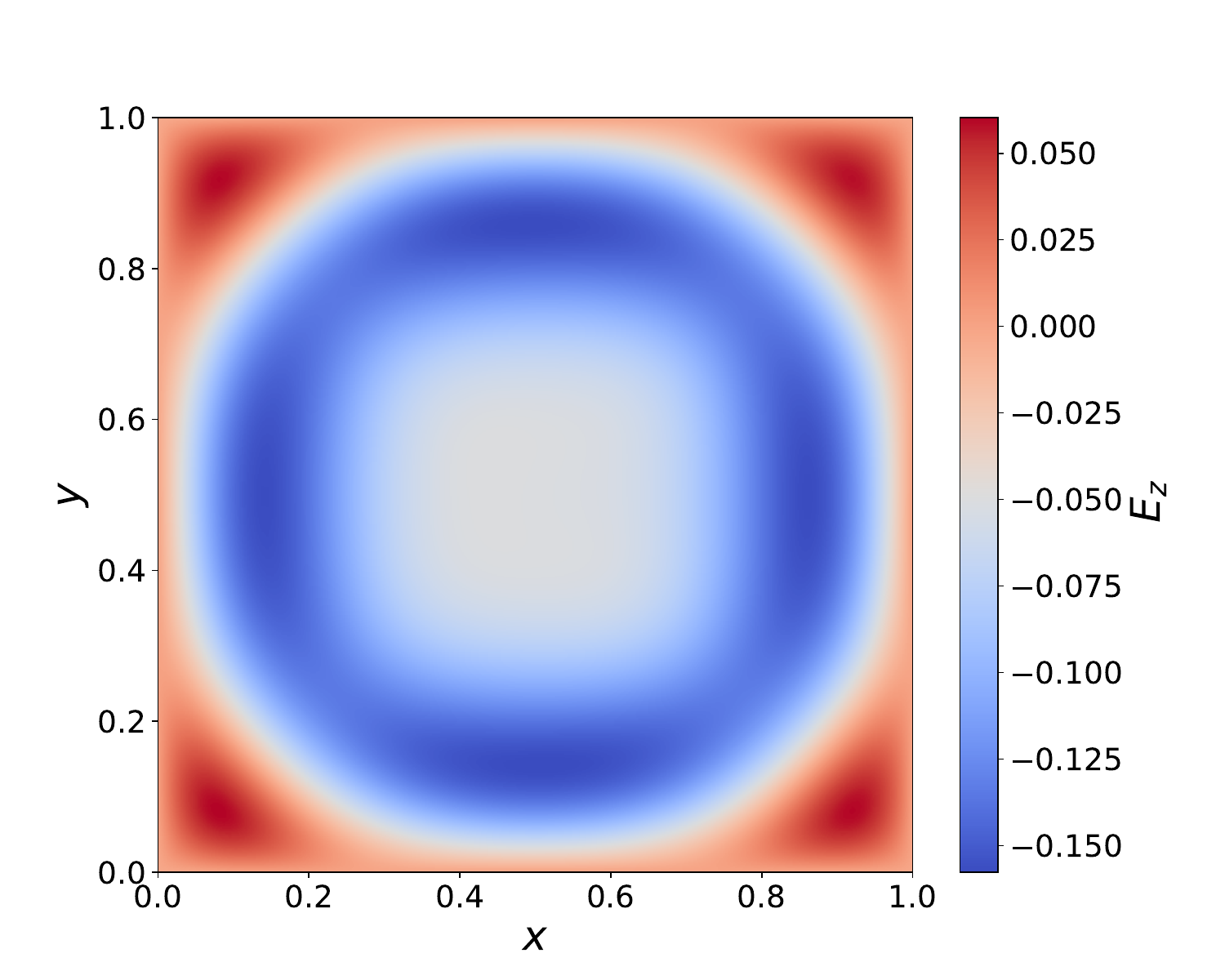}
  \caption{}
\end{subfigure}
\caption{Electric field $E_z$ at (a) t =0.4; (b)  t = 0.56. }
\label{fig:EzField2DGaussiant06t08}
\end{figure}

At $x = 0.5$ and $t = 0.2$, the PINNs solution closely matches the FDTD results in Fig.~\ref{fig:EzComp2DGaussianPML}.
The small discrepancy near the boundary arises because the FDTD simulation was performed on a sufficiently large domain to obtain an accurate reference solution, without employing a PML.
The $E_z$ fields at $t = 0.4$ and $t = 0.56$ are presented in Fig.~\ref{fig:EzField2DGaussiant06t08}. The outward-propagating pulse is effectively absorbed by the PML region.

\section{Conclusions and Future Work}
This paper explored split-field PML in the context of PINNs. The results show good agreement with reference solutions, demonstrating the potential of PINNs combined with PML for practical electromagnetic problems. Future work will focus on temporally discretized PINNs for solving higher-dimensional problems.



%

\end{document}